%% file: main.tex
\documentclass[10pt,journal,cspaper,compsoc]{IEEEtran}
\ifCLASSOPTIONcompsoc
\else
\fi
\usepackage{graphicx}

\usepackage[cmex10]{amsmath}
\usepackage{algpseudocode}
\ifCLASSOPTIONcompsoc
\usepackage[tight,normalsize,sf,SF]{subfigure}
\else
\usepackage[tight,footnotesize]{subfigure}
\fi
\usepackage{url}
\usepackage{acronym}
\usepackage{amssymb}
\usepackage{amsthm}
\usepackage{multirow}
\usepackage{tabularx}
\usepackage{supertabular}
\usepackage{rotating}
\usepackage{framed}
\usepackage{color}

\usepackage{float}
\floatstyle{ruled}
\newfloat{algorithm}{thp}{lop}
\floatname{algorithm}{Algorithm}

\begin{document}

\definecolor{gray}{gray}{.8} 

\newtheorem{thm}{Theorem}
\newtheorem{corollary}[thm]{Corollary}
\newtheorem{lemma}[thm]{Lemma}
\theoremstyle{definition}
\newtheorem{defi}[thm]{Definition}
\newtheorem{remark}[thm]{Remark}
\newenvironment{beweis}{\begin{proof}[Proof]}{\end{proof}}

\newcommand{\etal}{~\emph{et al.}}

\title{Correction of ``A Comparative Study to Benchmark Cross-project Defect
Prediction Approaches''}

%\author{Steffen~Herbold,~\IEEEmembership{Member,~IEEE}

\author{Steffen~Herbold, Alexander Trautsch, Jens Grabowski
\IEEEcompsocitemizethanks{\IEEEcompsocthanksitem S. Herbold, A. Trautsch, and J.
Grabowski are with the University of Goettingen, Institute of Computer Science,
G\"{o}ttingen, Germany.\protect\\
E-mail: \{herbold,grabowski\}@cs.uni-goettingen.de\protect\\
alexander.trautsch@stud.uni-goettingen.de}
% <-this
\thanks{}}

% \author{XXX}

% % stops a space


\markboth{IEEE Transactions on Software Engineering}%
{Shell \MakeLowercase{\textit{et al.}}: Bare Demo of IEEEtran.cls for Computer Society Journals}

%\IEEEpubid{\makebox[\columnwidth]{\hfill 0000--0000/00/\$00.00~\copyright~2007 IEEE}%

\input{acronyms}
\input{abstract}

\maketitle

% is not selected <OR> if conference mode is selected - because compsoc
% conference papers position the abstract like regular (non-compsoc)
\IEEEdisplaynotcompsoctitleabstractindextext

% \fi
\IEEEpeerreviewmaketitle

\acresetall
\input{introduction}
\input{problem}

\input{results}
\input{conclusion}


% \ifCLASSOPTIONcompsoc
%   % The Computer Society usually uses the plural form
%   \section*{Acknowledgments}
% \else
%   % regular IEEE prefers the singular form
% \fi

\ifCLASSOPTIONcaptionsoff
  \newpage
\fi



\bibliographystyle{IEEEtran}
\bibliography{literature}

\end{document}

%% file: acronyms.tex
\acrodef{ANOVA}{ANalysis Of VAriance}
\acrodef{AST}{Abstract Syntax Tree}
\acrodef{AUC}{Area Under the ROC Curve}
\acrodef{Ca}{Afferent Coupling}
\acrodef{CBO}{Coupling Between Objects}
\acrodef{CCA}{Canonical Correlation Analysis}
\acrodef{CD}{Critical Distance}
\acrodef{Ce}{Efferent Coupling}
\acrodef{CFS}{Correlation-based Feature Subset}
\acrodef{CLA}{Clustering and LAbeling}
\acrodef{CODEP}{COmbined DEfect Predictor}
\acrodef{CPDP}{Cross-Project Defect Prediction}
\acrodef{DBSCAN}{Density-Based Spatial Clustering}
\acrodef{DCV}{Dataset Characteristic Vector}
\acrodef{DTB}{Double Transfer Boosting}
\acrodef{fn}{false negative}
\acrodef{fp}{false positive}
\acrodef{GB}{GigaByte}
\acrodef{HL}{Hosmer-Lemeshow}
\acrodef{ITS}{Issue Tracking System}
\acrodef{JIT}{Just In Time}
\acrodef{LCOM}{Lack of COhession between Methods}
\acrodef{LOC}{Lines Of Code}
\acrodef{MDP}{Metrics Data Program}
\acrodef{MI}{Metric and Instances selection}
\acrodef{MODEP}{MultiObjective DEfect Predictor}
\acrodef{MPDP}{Mixed-Project Defect Prediction}
\acrodef{NN}{Nearest Neighbor}
\acrodef{PCA}{Principle Component Analysis}
\acrodef{RAM}{Random Access Memory}
\acrodef{RFC}{Response For a Class}
\acrodef{SCM}{SourceCode Management system}
\acrodef{SVM}{Support Vector Machine}
\acrodef{TCA}{Transfer Component Analysis}
\acrodef{tn}{true negative}
\acrodef{tp}{true positive}
\acrodef{RBF}{Radial Basis Function}
\acrodef{ROC}{Receiver Operating Characteristic}
\acrodef{UMR}{Unified Metric Representation}
\acrodef{VCB}{Value-Cognitive Boosting}
\acrodef{WPDP}{Within-Project Defect Prediction}

%% file: abstract.tex
% PRIOR to the title within the \IEEEcompsoctitleabstractindextext IEEEtran
\IEEEcompsoctitleabstractindextext{%
\begin{abstract}
%\boldmath
Unfortunately, the article ``A Comparative Study to Benchmark Cross-project Defect
Prediction Approaches'' has a problem in the statistical analysis which was
pointed out almost immediately after the pre-print of the article appeared
online. While the problem does not negate the contribution of the the article
and all key findings remain the same, it does alter some rankings of approaches
used in the study. Within this correction, we will explain the problem, how we
resolved it, and present the updated results. 
\end{abstract}
% not want either math or citations to appear in the abstract.

% Note that keywords are not normally used for peer review papers.
}

%% file: introduction.tex
\section{Introduction}
\label{sec:introduction}

Unfortunately, the article ``A Comparative Study to Benchmark Cross-project
Defect Prediction Approaches''~\cite{Herbold2017b} has a problem in the
statistical analysis performed to rank \ac{CPDP} approaches.
Prof. Yuming Zhou from Nanjing University
pointed out an inconsistency in Table 8 of the the article. He noted that in
some cases the $rankscores$ are worse even if the mean values for
performance metrics are better. While this is possible in theory, with the Friedman
test~\cite{Friedman1940} with post-hoc Nemenyi test~\cite{Nemenyi1963}, such inconsistencies are unlikely.
Therefore, we immediately proceeded to check our results. These checks revealed that the inconsistencies are due to a problem
with our statistical analysis for the Research Question 1 (RQ1) ``Which
CPDP approaches perform best in terms of F-measure, G-measure, AUC, and MCC?''. None
of the raw results of the benchmark, nor any of the other research questions are
affected by the problem.

We will describe the problem and how we solved in in Section~\ref{sec:problem}.
Then, we will show the updated results regarding RQ1 and discuss the changes in
Section~\ref{sec:results}. Afterwards, we analyze the reasons for the changes in
Section~\ref{sec:reasons} to determine if all changes due to the correction are
plausible and the correction resolves the inconsistencies reported by Y. Zhou.
In Section~\ref{sec:replication}, we describe how we updated our replication kit
as part of this correction. Finally, we will conclude in
Section~\ref{sec:conclusion}.
Please note, that we assume that readers have read to the original article and are
familiar with the terminology used. We do not re-introduce any of the
terminology in this correction. 

%% file: problem.tex
\section{Problem with the Nemenyi test implementation}
\label{sec:problem}

On July 15th 2017, Y. Zhou imformed us that he found an inconsistency between
the results of CV-NET and CamargoCruz09-DT for the RELINK data for the performance
metric AUC. He noted that the mean value for CV-NET was higher than for
CamargoCruz09-DT, but the $rankscore$ was lower. He went to the raw data
provided as part of the replication kit~\cite{Herbold2017a} and confirmed that
the mean values were correct, and the AUC for CV-NET was higher for all three
products of the RELINK data. Based on this observervation, we re-checked our
statistical analysis of the results. We found the problem in our implementation
of the Nemenyi post-hoc test.

\subsection{Summary of the Friedman and Nemenyi tests}

To understand the problem, we briefly recap how the Friedman test with
post-hoc Nemenyi test works. The Friedman test determines if there are
stastical significant differences between populations. This is done using
pair-wise comparisons between the rankings of populations. If the Friedman test
determines significant differences, the Nemenyi post-hoc test compares the
populations to each other to determine the statistically significantly
different ones. The analysis with the Nemenyi test is based on two parameters:
the \ac{CD} and the average ranks of the populations in the pair-wise comparisons
between all populations on each data set.
Following the description by Dem\v{s}ar~\cite{Demsar2006}, \ac{CD} is defined as
\begin{equation}
CD = q_\alpha\sqrt{\frac{k(k+1)}{6N}}
\end{equation}
where $q_\alpha = \frac{qtukey(\alpha, N, \inf)}{\sqrt{2}}$ is the studentized
range distribution with infinite degrees of freedom divided\footnote{For simplicity, we refer to the studentized range
distribution as $qtukey(\alpha, N)$, following the name of the related method in
R} by $\sqrt{2}$, $\alpha$ the significance level, $k$ the number of populations
compared and $N$ the number of data sets. We can thus rewrite \ac{CD} as
\begin{equation}
\begin{split}
CD &= \frac{qtukey(\alpha, N, \inf)}{\sqrt{2}}\sqrt{\frac{k(k+1)}{6N}} \\ 
&= qtukey(\alpha, N, \inf)\sqrt{\frac{k(k+1)}{12N}}
\end{split}
\end{equation}

If we now assume that $R_i, R_j$ are the average ranks of population 
$i,j \in \{1,\ldots,N\}$, the two populations are stastically significantly
different if
\begin{equation}
|R_i-R_j| > CD.
\end{equation}

In case a control population is available, it is possible to use a procedure
like Bonferroni correction~\cite{Dunn1961}. In this case, all populations are compared
to the control classifier instead of to each other. This greatly reduces the
number of pair-wise comparisons and can make the test more powerful. In this
case, for each pair a $z$-value is computed as
\begin{equation}
z = (R_i-R_j)/\sqrt{\frac{k(k+1)}{6N}}. 
\end{equation}
The $z$-values are then used to rank classifiers. Since we do not have a control classifier and have to do pair-wise
comparisons with the \ac{CD} and cannot make use of the $z$-values. However, the
$z$-values play an important role when it comes to the problem with our analysis. 

\subsection{z-values instead of Ranks}

Now that the concepts of the statistical tests are introduced, we can discuss
the actual problem in our implementation. We used the
\texttt{posthoc.friedman.nemenyi.test} function of the PMCMR
package~\cite{Pohlert2014} to implement the test. As part of the return values,
the function returns a matrix called \texttt{PSTAT}. Without checking directly
in the source code, we assumed these were the average ranks for each population, based on the documention of the package.
However, these are actually the absolute $z$-values multiplied with $\sqrt{2}$,
i.e., 
\begin{equation}
z' = |R_i-R_j|/\sqrt{\frac{k(k+1)}{6N}\cdot\sqrt{2}}. 
\end{equation}
Thus, when we compared ranks, we did not actually compare the average ranks, but the
mean $z'$-values. This led to a wrong determination of ranks, which explain the
inconsistencies found by Y. Zhou.

\subsection{The solution}
To resolve the problem, we adopted the code from the PMCMR package that
determines the average ranks. We cross-checked our code with another
implementation of the Nemenyi test~\cite{Svetunkov2017}, to ensure that
the new code solves the problem.\footnote{Both implementations of the test do not return the raw
pair-wise comparisons and can, therefore, not be used directly.} We then used
the mean averages from that code, instead of the $z$-values that were returned
from the PMCMR package. As a result, the Nemenyi-test became much more
sensitive, because the scale of the average ranks is different than the scale of the $z$-values. Let us consider the scales for our experiments
with the JURECZKO data. Here, we have $N=62$ data sets and $k=135$ populations,
i.e., \ac{CPDP} approach and classifier combinations. The best possible average
rank is 135 (always wins), the worst possible is 1 (always loses). Thus, the
average ranks are on a scale from 1 to 135. In comparison, the highest possible
$z'$-value is
\begin{equation}
z = (135-1)/\sqrt{\frac{135\cdot136}{6\cdot62}}\cdot\sqrt{2} \approx 26.97,
\end{equation}
i.e., the scale is just from 0 (no difference in average ranks) and 26.97. Thus,
the scale of the $z'$-values has only about a fifth of the range the scale of
the average ranks has. Basically, with $z'$-values, 135
populations are fit into the scale 0 to 26.97, with rankings in the scale 1 to 135. This means that the
average distance between approaches is 0.2 with $z'$-values and 1 in case of
average ranks. Considering that we have a $CD \approx 1.26$ for $\alpha=0.95$
in this example, this makes a huge difference. With $z'$-values, it is unlikely
that two subsequently ranked approaches to have
a distance greater than CD, because $CD$ was more than 6.3 times higher than
average distance expected on the scale. This changes if the real scale with ranks is used. If you
have 135 cases with an average distance of 1, it is quite likely that a few of
these distances will be greater than 1.26, i.e., the CD. 

We discuss this change in scales in this detail, because it requires a small
change in the ranking of approaches based on the Nemenyi test. Before, we considered three distinct ranks for
the calculation of the rankscore to achieve non-overlapping ranks:
\begin{itemize}
  \item The populations that are within the CD of the best average ranking
  population (top rank 1).
  \item The populations that are within the CD of the worst average ranking
  population (bottom rank~3).
  \item The populations that are neither (middle rank 2). 
\end{itemize}

This was the only way to deal with the small differences that resulted from
using the $z'$-values. However, this approach breaks on the larger scale,
because the distances now become larger, meaning fewer results are within the CD from
the best/worst ranking. For example, for the JURECZKO data and the performance
metric AUC, only two approaches would be on the first rank, i.e., only one
approach is within the CD of the best approach. Similarly, only six approaches
would be on the third rank, i.e., only five approaches are within the CD of the
worst approach. This would leave us with 127 approaches on the middle rank. This
ranking would be to coarse, and not show actual differences between approaches
anymore. To deal with this larger scale of ranks, we use a simple and more
fine-grained grouping strategy to create the ranks. We sort all approaches by their average ranking.
Each time that the difference between two subsequently ranked approaches is
larger than the \ac{CD}, we increase the rank. Because the rank is only
increased if the difference is larger than the \ac{CD}, we ensure that each
group only contains approaches that are statistically significantly different from the
other groups. Afterwards, we calculate the normalized rankscore as before.
Algorithm~\ref{alg:ranking} formalizes this strategy. This change in ranking
increases the sensitivity of the test and makes the results more fine-grained in
comparison to our original ranking procedure. 

\begin{algorithm}
\caption{Ranking algorithm.}
\label{alg:ranking}
\label{alg:abstract_learner}
\begin{algorithmic}[1]
\State \textbf{Input:} Sorted mean ranks $R_i$ such that $\forall~i,j \in
\{1,\ldots,N\}:i<j, R_i\geq R_j$
\State \textbf{Output:} $rankscore_i$ for all ranks.
\State $rank\_tmp_1 = 1$
\State $current\_rank \leftarrow 0$
\For{i=2,\ldots, N}
  \State \Comment{If difference is larger than CD increase rank}
  \If{$R_{i-1}-R_i>CD$}
    \State $current\_rank \leftarrow current\_rank+1$
  \EndIf
  \State $rank\_tmp_i \leftarrow current\_rank$
\EndFor
\State $rank\_max \leftarrow current\_rank$
\State \Comment{Determine rankscores}
\For{i=1,\ldots, N}
  \State $rankscore_i \leftarrow 1-\frac{rank\_tmp_i}{rank\_max}$
\EndFor
\end{algorithmic}
\end{algorithm}

%% file: results.tex
\section{Results}
\label{sec:results}

\begin{figure}
\includegraphics[width=\linewidth]{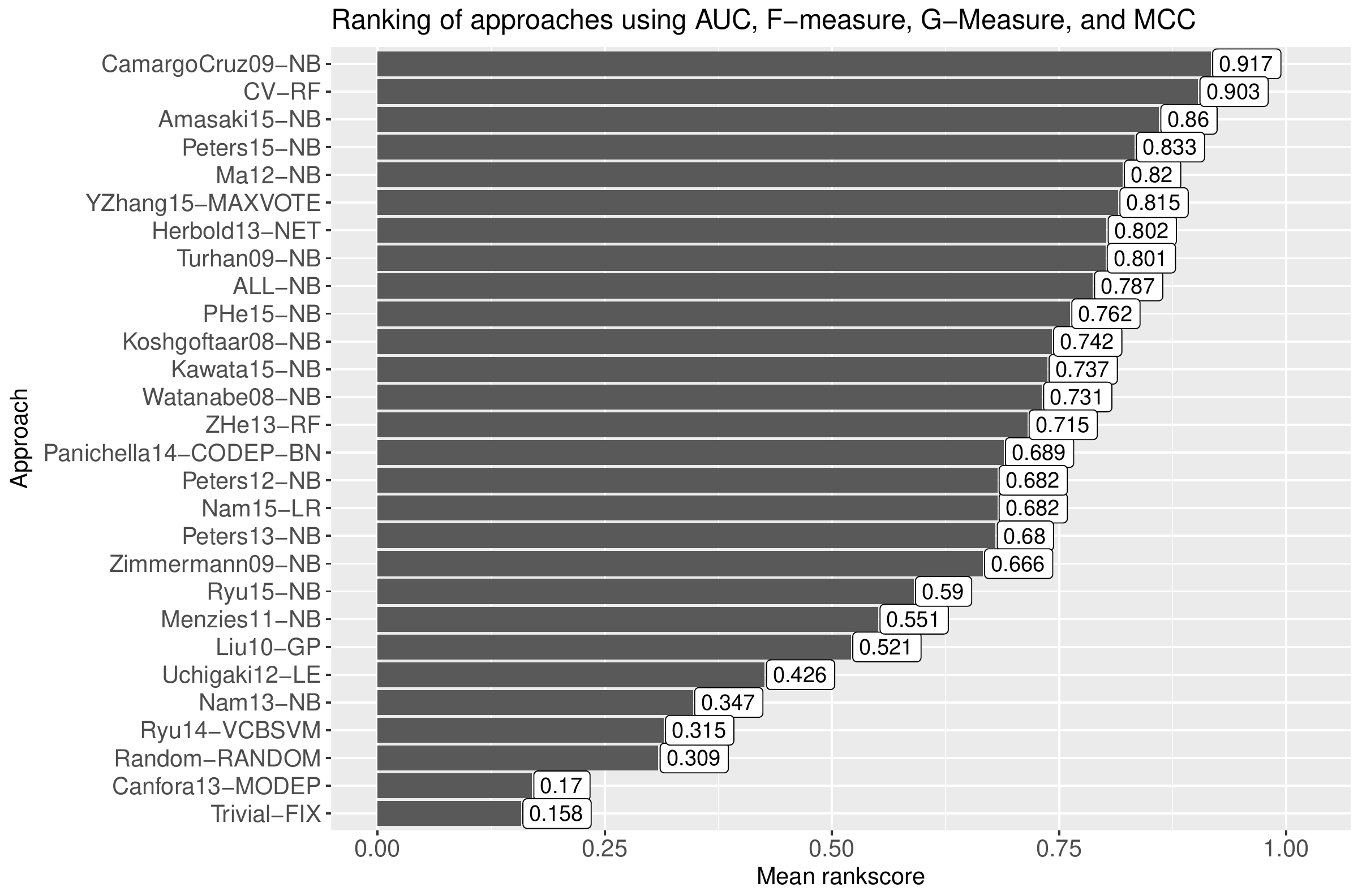}
\caption{Mean rank score over all data sets for the metrics \emph{AUC},
\emph{F-measure}, \emph{G-measure}, and \emph{MCC}. In case multiple classifiers
were used, we list only the result achieved with the best classifier.}
\label{fig:best-rq1}
\end{figure}

\input{resultsTable.tex}

We now show the corrected results for RQ1. We will directly
compare the changes in the results with the originally published results.
Figure~\ref{fig:best-rq1} shows the mean \emph{rankscore} averaged over the four
performance metrics F-Measure, G-Measure, AUC, and MCC and the five data sets
JURECZKO, MDP, AEEEM, NETGENE, and RELINK. Table~\ref{tbl:results} shows
detailed results including the mean values and rankscores for each performance
metrics and each data set. Figure~\ref{fig:best-rq1} is the correction of
Figure~3 and Table~\ref{tbl:results} the correction of Table~8 from the original
publication. Table~\ref{tbl:results} and Figure~\ref{fig:best-rq1} only report
the results for the best classifier for each approach. In case these changed between the original results
and our correction, you will not find the exact same rows. For example, for
CamargoCruz09, we reported DT as best classifier in the original analysis, and
now NB. This is because with the problem in the statistical analysis DT was
ranked best for CamargoCruz09, but in the corrected version NB performs better.
The reasons for these, and other changes are explained in
Section~\ref{sec:reasons}

The most important finding remains the same: the
approach CamargoCruz09 still provides the best-ranking classification model with
a mean \emph{rankscore} of 0.917 for CamargoCruz09-NB. However, the
\emph{rankscore} is not a perfect 1.0 anymore. We attribute this to the more
sensitive ranking due to the correction of the Nemenyi test.
The differences to the next-ranking approaches are still rather small, though the
group of approaches that is within 10\% of the best ranking approach now only consists of CV-RF, Amasaki15, Peters15, and Ma12. The bottom of the ranking is nearly unaffected by the changes as well. The last
seven ranked approaches are still the same. Additionally, our findings regarding
the comparison of using ALL data, versus transfer learning approaches has not
changed: ALL is still in the upper mid-field of approaches. With the corrected
and more finegrained ranking, only six of the cross-project approaches actually
outperform this baseline, whereas seventeen are actually worse.

With respect to CV versus \ac{CPDP}, we still observe that \ac{CPDP} can
outperform CV in case multiple performance criteria are considered because
CV-RF is outperformed by CamargoCruz09-NB. Thus, we still note that this is
possible, but far less conclusively than before, where CV was actually only in
the mid-field of the approaches and not a close second.

Due to these small overall small differences, we change our answer to RQ1
slightly: 
\begin{framed}
\noindent\textbf{Answer RQ1:}
CamargoCruz09-NB performs best among the compared \ac{CPDP} approaches and
even outperforms cross-validation. However, the differences to other
approaches are small. The baseline ALL-NB is ranked higher than seventeen of the
\ac{CPDP} approaches.
\end{framed}

\section{Reasons for changes}
\label{sec:reasons}

We checked our raw results for the reasons for all changes in rankings. The
problem with the statistical analysis actually led to two reasons for ranking
changes: first, the $z$-values already consider differences in ranks. Thus, if
the rank was very high, this could would lead to larger $z$-values, which would
negatively impact the ranking. Second, because differences were downscaled with
the $z$-values in comparison to differences in mean ranks, too many approaches
were grouped together as not statistically significantly different. For
approaches that are now better ranked than before, this means that they were
often among the best performing approaches within a group. For those that are
now ranking worse, they were at often near the bottom of their groups. For
example, CV was often among the best approaches on the middle rank. Now, it is
clearly distinguished from the others there, leading to the strong rise in the
ranking. Others that were affected the same way, though to a lesser extend are
Amasaki15, Peters15, YZhang15, and Herbold13. On the other hand Menzies11 and
Watanabe08 were often at the bottom of their groups, leading to the big losses
in rankings for both approaches.

Another change in our results is that NB is often the best
performing classifier, whereas before DT and LR were most often the best
performing classifiers. We previously already noted in our discussion that ``for
many approaches the differences between the classifiers were rather
small''~\cite{Herbold2017b}. Together with the reasons for ranking
changes explained above, theses changes are not unexpected. 

Overall, all changes in the result are plausible. Moreover, our comparison
of the results of the statistical analysis with both mean values, as well as the
raw results of the benchmark did not reveal any inconsistencies of the type that
Y. Zhou reported to us. Therefore, we believe that the problem was
correctly resolved. 

\section{Update of the replication kit}
\label{sec:replication}

We updated the replication kit archived at Zenodo~\cite{Herbold2017c}. The
changes two the replication kit are two-fold.
\begin{itemize}
  \item We corrected the problem with the statistical analysis in the
  generate\_results.R script.
  \item We updated the provided CD diagrams due to the changes in the Nemenyi
  test.
\end{itemize}

The changes can be reviewed in detail in the commit to the GitHub archive of the
replication kit\footnote{https://goo.gl/AbvSRj}. 

%% file: resultsTable.tex
% latex table generated in R 3.3.3 by xtable 1.8-2 package
% Fri Jul 21 09:23:06 2017
\begin{table*}
\scriptsize\centering\begin{sideways}

\begin{tabular}{rllllllllllll}
  \hline
 & \multicolumn{4}{c|}{JURECZKO} & \multicolumn{4}{c|}{FILTERJURECZKO / SELECTEDJURECZKO} & \multicolumn{4}{c|}{MDP} \\
\hline
 & \emph{AUC} & \emph{F-measure} & \emph{G-measure} & \emph{MCC} & \emph{AUC} & \emph{F-measure} & \emph{G-measure} & \emph{MCC} & \emph{AUC} & \emph{F-measure} & \emph{G-measure} & \emph{MCC} \\ 
  \hline
ALL-NB & \textbf{0.72 (1)} & 0.28 (0.6) & 0.35 (0.48) & 0.2 (0.53) & -0.02 / 0.02 & 0.02 / 0.08 & 0.03 / 0.08 & 0.00 / 0.05 & 0.75 (0.88) & 0.26 (0.88) & 0.41 (0.85) & 0.21 (0.92) \\ 
  Amasaki15-NB & \textbf{0.73 (1)} & 0.47 (0.95) & 0.61 (0.9) & 0.28 (0.76) & -0.02 / 0.01 & 0.00 / 0.04 & 0.02 / 0.04 & -0.02 / 0.02 & \textbf{0.77 (1)} & 0.31 (0.93) & 0.63 (0.97) & 0.22 (0.97) \\ 
  CamargoCruz09-NB & \textbf{0.73 (1)} & 0.47 (0.95) & 0.62 (0.95) & 0.27 (0.76) & -0.02 / 0.01 & 0.00 / 0.04 & 0.00 / 0.03 & -0.02 / 0.02 & 0.77 (0.98) & 0.28 (0.86) & 0.58 (0.94) & 0.22 (0.95) \\ 
  Canfora13-MODEP & 0.52 (0.12) & 0.44 (0.75) & 0.48 (0.67) & 0.19 (0.53) & 0.01 / 0.03 & 0.04 / 0.09 & 0.04 / 0.14 & 0.00 / 0.08 & 0.5 (0.02) & 0.06 (0.26) & 0.04 (0.18) & -0.01 (0) \\ 
  CV-RF & \textbf{0.76 (1)} & 0.51 (0.9) & 0.52 (0.67) & 0.34 (0.88) & 0.00 / 0.03 & 0.01 / 0.03 & 0.02 / 0.06 & 0.01 / 0.05 & 0.76 (0.91) & 0.31 (0.81) & 0.35 (0.74) & 0.28 (0.92) \\ 
  Herbold13-NET & \textbf{0.71 (1)} & 0.48 (0.9) & 0.54 (0.71) & 0.22 (0.53) & -0.01 / 0.02 & -0.01 / 0.01 & 0.04 / 0.08 & 0.00 / 0.03 & 0.75 (0.8) & 0.25 (0.78) & 0.45 (0.82) & 0.17 (0.75) \\ 
  Kawata15-NB & \textbf{0.71 (1)} & 0.28 (0.6) & 0.34 (0.43) & 0.2 (0.53) & -0.02 / 0.02 & 0.02 / 0.09 & 0.03 / 0.09 & 0.00 / 0.06 & 0.72 (0.73) & 0.22 (0.72) & 0.32 (0.73) & 0.19 (0.87) \\ 
  Koshgoftaar08-NB & 0.63 (0.56) & 0.39 (0.85) & 0.5 (0.71) & 0.26 (0.76) & 0.00 / 0.00 & 0.00 / 0.03 & 0.00 / 0.01 & -0.01 / 0.02 & 0.62 (0.38) & 0.29 (0.91) & 0.47 (0.88) & 0.21 (0.93) \\ 
  Liu10-GP & 0.63 (0.5) & 0.51 (0.9) & 0.52 (0.67) & 0.23 (0.53) & -0.05 / -0.08 & -0.02 / -0.02 & -0.13 / -0.29 & -0.07 / -0.09 & 0.65 (0.45) & 0.27 (0.84) & 0.52 (0.89) & 0.17 (0.78) \\ 
  Ma12-NB & \textbf{0.72 (1)} & 0.34 (0.65) & 0.43 (0.62) & 0.24 (0.59) & -0.02 / 0.02 & 0.01 / 0.05 & 0.02 / 0.04 & -0.01 / 0.03 & 0.75 (0.88) & 0.31 (0.97) & 0.5 (0.92) & 0.24 (0.98) \\ 
  Menzies11-NB & 0.59 (0.31) & 0.34 (0.65) & 0.44 (0.62) & 0.19 (0.53) & -0.01 / 0.02 & 0.01 / 0.06 & 0.03 / 0.07 & -0.02 / 0.03 & 0.55 (0.23) & 0.19 (0.66) & 0.37 (0.79) & 0.08 (0.4) \\ 
  Nam13-NB & - & - & - & - & - & - & - & - & - & - & - & - \\ 
  Nam15-LR & 0.69 (0.81) & 0.51 (0.95) & 0.63 (0.9) & 0.29 (0.71) & -0.03 / 0.00 & -0.03 / 0.01 & -0.02 / 0.01 & -0.05 / -0.01 & 0.63 (0.39) & 0.26 (0.72) & 0.35 (0.7) & 0.13 (0.68) \\ 
  Panichella14-CODEP-BN & 0.63 (0.56) & 0.39 (0.75) & 0.51 (0.71) & 0.25 (0.65) & 0.00 / 0.02 & 0.01 / 0.06 & 0.02 / 0.06 & -0.01 / 0.03 & 0.55 (0.27) & 0.16 (0.64) & 0.2 (0.64) & 0.15 (0.72) \\ 
  Peters12-NB & \textbf{0.71 (1)} & 0.2 (0.35) & 0.24 (0.29) & 0.15 (0.53) & -0.02 / 0.01 & 0.02 / 0.08 & 0.02 / 0.08 & 0.00 / 0.07 & 0.73 (0.79) & 0.21 (0.71) & 0.31 (0.73) & 0.18 (0.83) \\ 
  Peters13-NB & \textbf{0.71 (1)} & 0.2 (0.35) & 0.24 (0.29) & 0.15 (0.53) & -0.02 / 0.01 & 0.02 / 0.08 & 0.02 / 0.08 & 0.00 / 0.07 & 0.73 (0.79) & 0.22 (0.71) & 0.31 (0.73) & 0.18 (0.85) \\ 
  Peters15-NB & \textbf{0.71 (1)} & 0.47 (0.95) & 0.61 (0.9) & 0.26 (0.71) & 0.00 / 0.04 & 0.01 / 0.04 & 0.01 / 0.04 & 0.00 / 0.02 & 0.77 (0.98) & 0.34 (0.98) & 0.63 (0.98) & \textbf{0.25 (1)} \\ 
  PHe15-NB & \textbf{0.74 (1)} & 0.46 (0.95) & 0.6 (0.9) & 0.3 (0.88) & -0.04 / -0.01 & -0.04 / 0.02 & -0.03 / 0.01 & -0.06 / -0.02 & 0.72 (0.73) & 0.24 (0.81) & 0.47 (0.85) & 0.18 (0.8) \\ 
  Random-RANDOM & 0.5 (0) & 0.37 (0.65) & 0.49 (0.67) & 0.00 (0) & 0.00 / 0.00 & 0.00 / 0.00 & 0.01 / 0.01 & 0.00 / 0.00 & 0.51 (0.09) & 0.18 (0.53) & 0.5 (0.91) & 0.01 (0.15) \\ 
  Ryu14-VCBSVM & 0.6 (0.38) & 0.46 (0.75) & 0.5 (0.67) & 0.18 (0.53) & -0.01 / 0.03 & -0.02 / 0.03 & 0.02 / 0.09 & -0.02 / 0.05 & 0.56 (0.2) & 0.24 (0.71) & 0.22 (0.38) & 0.07 (0.4) \\ 
  Ryu15-NB & 0.62 (0.56) & 0.44 (0.8) & 0.58 (0.81) & 0.22 (0.53) & -0.01 / -0.01 & -0.01 / -0.01 & 0.00 / 0.00 & -0.02 / -0.02 & 0.64 (0.43) & 0.29 (0.9) & 0.6 (0.95) & 0.18 (0.8) \\ 
  Trivial-FIX & 0.5 (0) & 0.48 (0.7) & 0.00 (0) & 0.00 (0) & 0.00 / 0.00 & -0.01 / -0.01 & 0.00 / 0.00 & 0.00 / 0.00 & 0.5 (0.04) & 0.21 (0.59) & 0.00 (0) & 0.00 (0.05) \\ 
  Turhan09-NB & \textbf{0.73 (1)} & \textbf{0.5 (1)} & \textbf{0.64 (1)} & 0.29 (0.88) & -0.02 / 0.01 & -0.01 / 0.02 & -0.01 / 0.01 & -0.02 / 0.00 & 0.77 (0.96) & \textbf{0.34 (1)} & \textbf{0.65 (1)} & \textbf{0.25 (1)} \\ 
  Uchigaki12-LE & \textbf{0.74 (1)} & 0.08 (0.15) & 0.09 (0.1) & 0.1 (0.29) & -0.03 / 0.01 & 0.01 / 0.11 & 0.01 / 0.12 & 0.02 / 0.11 & 0.77 (0.98) & 0.07 (0.31) & 0.08 (0.33) & 0.09 (0.47) \\ 
  Watanabe08-NB & \textbf{0.71 (1)} & 0.32 (0.65) & 0.4 (0.57) & 0.21 (0.53) & -0.02 / 0.03 & 0.00 / 0.04 & 0.00 / 0.03 & -0.01 / 0.04 & 0.73 (0.73) & 0.14 (0.55) & 0.23 (0.61) & 0.14 (0.68) \\ 
  YZhang15-MAXVOTE & \textbf{0.74 (1)} & 0.45 (0.95) & 0.58 (0.86) & 0.29 (0.88) & -0.02 / 0.01 & 0.00 / 0.05 & 0.01 / 0.05 & -0.02 / 0.02 & 0.76 (0.88) & 0.31 (0.95) & 0.61 (0.95) & 0.23 (0.97) \\ 
  ZHe13-RF & 0.65 (0.62) & 0.48 (0.95) & 0.56 (0.76) & 0.27 (0.76) & -0.02 / 0.00 & -0.01 / 0.00 & 0.00 / 0.04 & -0.05 / -0.01 & 0.63 (0.41) & 0.25 (0.81) & 0.52 (0.88) & 0.17 (0.78) \\ 
  Zimmermann09-NB & 0.66 (0.62) & 0.42 (0.7) & 0.52 (0.67) & 0.24 (0.59) & -0.02 / 0.04 & -0.06 / 0.01 & -0.06 / -0.01 & -0.07 / 0.04 & 0.72 (0.75) & 0.23 (0.72) & 0.35 (0.77) & 0.18 (0.77) \\ 
   \hline
\hline
 & \multicolumn{4}{c|}{AEEEM} & \multicolumn{4}{c|}{NETGENE} & \multicolumn{4}{c|}{RELINK} \\
\hline & \emph{AUC} & \emph{F-measure} & \emph{G-measure} & \emph{MCC} & \emph{AUC} & \emph{F-measure} & \emph{G-measure} & \emph{MCC} & \emph{AUC} & \emph{F-measure} & \emph{G-measure} & \emph{MCC} \\ 
  \hline
ALL-NB & 0.72 (0.69) & 0.37 (0.74) & 0.52 (0.81) & 0.26 (0.65) & 0.63 (0.59) & 0.31 (0.82) & 0.56 (0.86) & 0.16 (0.55) & 0.79 (0.96) & 0.67 (0.99) & 0.68 (0.99) & 0.45 (0.98) \\ 
  Amasaki15-NB & 0.74 (0.8) & 0.41 (0.9) & 0.61 (0.93) & 0.29 (0.77) & 0.63 (0.61) & 0.34 (0.79) & 0.5 (0.74) & 0.19 (0.59) & 0.74 (0.88) & 0.63 (0.95) & 0.63 (0.91) & 0.38 (0.87) \\ 
  CamargoCruz09-NB & 0.77 (0.96) & 0.44 (0.97) & \textbf{0.64 (1)} & 0.32 (0.9) & 0.68 (0.82) & 0.37 (0.9) & 0.6 (0.96) & 0.26 (0.81) & 0.74 (0.88) & 0.61 (0.93) & 0.64 (0.94) & 0.38 (0.88) \\ 
  Canfora13-MODEP & 0.49 (0) & 0.16 (0.26) & 0.18 (0.21) & 0.00 (0) & 0.5 (0.08) & 0.00 (0) & 0.00 (0) & 0.00 (0.08) & 0.5 (0.02) & 0.15 (0.14) & 0.1 (0.07) & 0.00 (0.01) \\ 
  CV-RF & 0.79 (0.98) & 0.41 (0.77) & 0.44 (0.57) & \textbf{0.36 (1)} & \textbf{0.86 (1)} & \textbf{0.54 (1)} & 0.59 (0.97) & \textbf{0.51 (1)} & \textbf{0.83 (1)} & 0.63 (0.98) & 0.66 (0.98) & \textbf{0.47 (1)} \\ 
  Herbold13-NET & 0.73 (0.81) & 0.42 (0.88) & \textbf{0.66 (1)} & 0.29 (0.78) & 0.69 (0.83) & 0.35 (0.89) & 0.47 (0.68) & 0.22 (0.79) & 0.76 (0.95) & 0.49 (0.73) & 0.53 (0.74) & 0.32 (0.65) \\ 
  Kawata15-NB & 0.72 (0.69) & 0.37 (0.74) & 0.52 (0.81) & 0.26 (0.66) & 0.61 (0.5) & 0.29 (0.66) & 0.53 (0.87) & 0.14 (0.44) & 0.77 (0.92) & 0.64 (0.96) & 0.66 (0.93) & 0.43 (0.95) \\ 
  Koshgoftaar08-NB & 0.64 (0.45) & 0.38 (0.74) & 0.59 (0.87) & 0.24 (0.55) & 0.63 (0.64) & 0.36 (0.93) & 0.52 (0.83) & 0.24 (0.85) & 0.67 (0.53) & 0.58 (0.89) & 0.58 (0.87) & 0.34 (0.7) \\ 
  Liu10-GP & 0.6 (0.33) & 0.36 (0.45) & 0.41 (0.54) & 0.18 (0.32) & 0.53 (0.28) & 0.27 (0.65) & 0.18 (0.28) & 0.06 (0.37) & 0.56 (0.22) & 0.59 (0.84) & 0.29 (0.28) & 0.18 (0.31) \\ 
  Ma12-NB & 0.72 (0.69) & 0.4 (0.86) & 0.58 (0.92) & 0.28 (0.76) & 0.63 (0.57) & 0.32 (0.85) & 0.58 (0.91) & 0.18 (0.62) & 0.76 (0.91) & 0.61 (0.9) & 0.62 (0.88) & 0.39 (0.94) \\ 
  Menzies11-NB & 0.61 (0.38) & 0.35 (0.58) & 0.55 (0.8) & 0.2 (0.36) & 0.58 (0.49) & 0.29 (0.76) & 0.49 (0.77) & 0.12 (0.52) & 0.63 (0.41) & 0.51 (0.54) & 0.55 (0.66) & 0.27 (0.56) \\ 
  Nam13-NB & - & - & - & - & - & - & - & - & 0.69 (0.64) & 0.39 (0.18) & 0.4 (0.22) & 0.25 (0.35) \\ 
  Nam15-LR & 0.68 (0.57) & 0.41 (0.74) & 0.64 (0.9) & 0.26 (0.53) & 0.6 (0.53) & 0.25 (0.62) & 0.49 (0.8) & 0.12 (0.48) & 0.68 (0.59) & 0.56 (0.82) & 0.5 (0.57) & 0.32 (0.61) \\ 
  Panichella14-CODEP-BN & 0.64 (0.45) & 0.38 (0.88) & 0.54 (0.82) & 0.29 (0.83) & 0.61 (0.58) & 0.31 (0.77) & 0.49 (0.71) & 0.18 (0.67) & 0.68 (0.56) & 0.56 (0.73) & 0.58 (0.88) & 0.38 (0.94) \\ 
  Peters12-NB & 0.69 (0.61) & 0.32 (0.56) & 0.45 (0.65) & 0.24 (0.56) & 0.6 (0.41) & 0.29 (0.65) & 0.5 (0.74) & 0.12 (0.4) & 0.78 (0.97) & 0.64 (0.97) & 0.66 (0.95) & 0.44 (0.97) \\ 
  Peters13-NB & 0.69 (0.6) & 0.32 (0.56) & 0.45 (0.65) & 0.24 (0.57) & 0.6 (0.43) & 0.29 (0.63) & 0.5 (0.74) & 0.12 (0.38) & 0.79 (0.98) & 0.63 (0.94) & 0.65 (0.93) & 0.43 (0.96) \\ 
  Peters15-NB & 0.71 (0.69) & 0.39 (0.84) & 0.61 (0.96) & 0.26 (0.7) & 0.61 (0.54) & 0.22 (0.54) & 0.34 (0.62) & 0.18 (0.63) & 0.76 (0.93) & 0.6 (0.88) & 0.65 (0.92) & 0.37 (0.89) \\ 
  PHe15-NB & 0.75 (0.88) & 0.41 (0.96) & 0.55 (0.88) & 0.33 (0.95) & 0.59 (0.55) & 0.16 (0.28) & 0.2 (0.29) & 0.16 (0.55) & 0.76 (0.96) & 0.55 (0.78) & 0.53 (0.55) & 0.31 (0.67) \\ 
  Random-RANDOM & 0.51 (0.06) & 0.27 (0.3) & 0.51 (0.74) & 0.01 (0.03) & 0.5 (0.05) & 0.23 (0.31) & 0.5 (0.75) & 0.00 (0.07) & 0.5 (0) & 0.43 (0.39) & 0.49 (0.47) & 0.00 (0) \\ 
  Ryu14-VCBSVM & 0.52 (0.13) & 0.26 (0.34) & 0.1 (0.12) & 0.09 (0.12) & 0.5 (0.08) & 0.00 (0) & 0.00 (0) & 0.00 (0.08) & 0.6 (0.21) & 0.54 (0.68) & 0.42 (0.29) & 0.2 (0.25) \\ 
  Ryu15-NB & 0.53 (0.17) & 0.21 (0.29) & 0.33 (0.46) & 0.06 (0.1) & 0.59 (0.51) & 0.3 (0.76) & 0.55 (0.87) & 0.15 (0.58) & 0.61 (0.33) & 0.4 (0.65) & 0.44 (0.66) & 0.22 (0.63) \\ 
  Trivial-FIX & 0.5 (0.02) & 0.31 (0.35) & 0.00 (0) & 0.00 (0.02) & 0.5 (0.08) & 0.26 (0.48) & 0.00 (0) & 0.00 (0.08) & 0.5 (0.01) & 0.56 (0.72) & 0.00 (0) & 0.00 (0.01) \\ 
  Turhan09-NB & 0.72 (0.7) & 0.4 (0.87) & 0.62 (0.98) & 0.28 (0.72) & 0.58 (0.36) & 0.19 (0.42) & 0.3 (0.48) & 0.13 (0.45) & 0.74 (0.88) & 0.53 (0.77) & 0.59 (0.81) & 0.31 (0.73) \\ 
  Uchigaki12-LE & 0.77 (0.98) & 0.1 (0.14) & 0.1 (0.11) & 0.18 (0.31) & 0.71 (0.86) & 0.26 (0.48) & 0.00 (0) & 0.00 (0.08) & 0.75 (0.94) & 0.34 (0.23) & 0.34 (0.34) & 0.27 (0.43) \\ 
  Watanabe08-NB & 0.74 (0.86) & 0.41 (0.94) & 0.54 (0.79) & 0.32 (0.94) & 0.7 (0.84) & 0.27 (0.68) & 0.38 (0.59) & 0.2 (0.74) & 0.74 (0.86) & 0.51 (0.64) & 0.57 (0.65) & 0.33 (0.77) \\ 
  YZhang15-MAXVOTE & 0.75 (0.88) & 0.43 (0.95) & 0.63 (0.95) & 0.31 (0.89) & 0.66 (0.75) & 0.25 (0.59) & 0.4 (0.58) & 0.16 (0.64) & 0.74 (0.89) & 0.51 (0.51) & 0.51 (0.52) & 0.32 (0.7) \\ 
  ZHe13-RF & 0.64 (0.44) & 0.38 (0.64) & 0.6 (0.87) & 0.22 (0.43) & - & - & - & - & 0.66 (0.52) & 0.61 (0.91) & 0.63 (0.9) & 0.32 (0.75) \\ 
  Zimmermann09-NB & 0.71 (0.64) & 0.37 (0.83) & 0.55 (0.81) & 0.26 (0.66) & 0.64 (0.7) & 0.35 (0.92) & 0.47 (0.7) & 0.23 (0.81) & 0.72 (0.74) & 0.3 (0.29) & 0.29 (0.31) & 0.21 (0.32) \\ 
   \hline
\end{tabular}
\end{sideways}
\caption{Mean results over all products with rankscores in brackets. Bold-faced values are top-ranking for the metric on the data set. For FILTERJURECZKO and SELECTEDJURECKO, we show the difference in the mean values to JURECZKO.}
\label{tbl:results}
\end{table*}

%% file: conclusion.tex
\section{Conclusion}
\label{sec:conclusion}

A problem with the implementation of the Nemenyi post-hoc test led to incorrect
results being published in our benchmark paper on cross-project defect
prediction. The mistake only affected research question RQ1, the other three
research questions were not affected. Within this correction paper, we explained
the problem in the statistical test, how this problem affected our results,
presented the corrected, and explained the changes that occoured. The
major findings regarding RQ1 are not changed, including the best performing approach, the result that the na\"{i}ve
baseline of using all data outperforms most proposed transfer learning
approaches, as well as that cross-validation can be outperformed by \ac{CPDP}.
Thus, the contributions of the article are still valid. Still, the correction
leads to differences in the rankings which are properly corrected and
discussed here. We apologize for this mistake and hope that this timely
correction mitigates the potential negative impact the wrong results may have.

\section*{Acknowledgements}

We want to thank Yuming Zhou  from Nanjing University for pointing out the
inconsistencies in the results to us so fast, as well as the editors of this
journals who helped to determine how we should communicate this problem to the
community within days.